\documentclass[review,1p]{elsarticle}

\usepackage{hyperref}

\usepackage{graphicx}
\usepackage{amsmath}
\usepackage{amssymb}
\usepackage{multirow}
\usepackage{bbm}
\usepackage{comment}
\usepackage{color}

\journal{Pervasive and Mobile Computing}

\begin{document}

\begin{frontmatter}

\title{Joint Cell Selection and Resource Allocation Games with Backhaul Constraints}





\author[mymainaddress,mysecondaryaddress]{Jorge Ort\'in}

\author[mymainaddress]{Jos\'e Ram\'on G\'allego\corref{mycorrespondingauthor}}
\cortext[mycorrespondingauthor]{Corresponding author}
\ead{jrgalleg@unizar.es}

\author[mymainaddress]{Mar\'a Canales}

\address[mymainaddress]{Instituto de Investigaci\'on en Ingenier\'ia de Arag\'on, Universidad de Zaragoza, Spain}
\address[mysecondaryaddress]{Centro Universitario de la Defensa Zaragoza, Spain}




\begin{abstract}
In this work we study the problem of user association and resource allocation to maximize the proportional fairness of a wireless network with limited backhaul capacity. The optimal solution of this problem requires solving a mixed integer non-linear programming problem which generally cannot be solved in real time. We propose instead to model the problem as a potential game, which decreases dramatically the computational complexity and obtains a user association and resource allocation close to the optimal solution. Additionally, the use of a game-theoretic approach allows an efficient distribution of the computational burden among the computational resources of the network.
\end{abstract}

\begin{keyword}
cell selection; channel allocation; power control; game theory; potential games.
\end{keyword}

\end{frontmatter}

\section{Introduction}
\label{sec:intro}

The increasing use of wireless devices to connect to the Internet and the development of new multimedia services pose new challenges in the design of wireless networks. To meet this growing demand of wireless traffic, one promising approach both in Wi-Fi and cellular networks is a dense deployment of access nodes, each of them covering a small portion of space and with a high degree of overlapping coverage between them. In this scenario an ideal backhaul with unlimited capacity for each access node is neither realistic nor efficient from a cost perspective. Instead, existing copper and fiber infrastructure or low-cost wireless technology are expected to be the basis of a heterogeneous resource-constrained backhaul \cite{Alcatel-report}, \cite{NGMN-report}. Therefore, the backhaul constraints of the access nodes must be considered as well when performing resource management in the radio access \cite{Domenico-PIMRC}.

Additionally, in the last years there has been an increasing interest in moving part of the operation of the access nodes to a centralized computing equipment, as is the case with Cloud-RAN \cite{C-RAN-report} or centralized WLAN. The aim of this approach is two-fold: first, centralizing allows reducing the cost of the radio access nodes, and second, it also enables using more advanced signal processing and resource management algorithms \cite{Shi-ToWC}, \cite{Zhu-ToWC}. Although in general these approaches require a low-latency high-capacity backhaul, using centralized solutions is also possible in heterogeneous backhaul scenarios with limited capacity \cite{Bartelt-VTC}. In this case, the decision on the functionalities to be centralized is flexible and depends on the features of the backhaul network and the computing servers.

One of the main issues in these densely deployed scenarios is the problem of joint cell selection and radio resource management. Although these tasks have been performed typically separately, some recent works have shown that tackling them jointly can improve the network efficiency \cite{Torregoza-EURASIP}, \cite{Li-ToVT}. Nevertheless, this approach generally leads to a complex mixed integer non-lineal programming (MINLP) problem that can only be solved exactly for small scenarios. For more realistic scenarios, approximate algorithms must be used instead. 

In this context, we propose two potential games that address the problem of joint channel assignment, power allocation and cell selection in a generic and technology-agnostic wireless network. Specifically, our goal is to maximize a modified version of the proportional fairness of the system \cite{Kelly-ETT}. In our  scenario, we take into account both the radio and the backhaul restrictions of the access nodes. These restrictions are imposed by the presence of interference due to spectral reuse and by the limited capacity of the backhaul expected in future dense scenarios. We assume also the presence of a cloud computing platform that allows moving part of the radio resource management functionalities from the access nodes.

Potential games \cite{Monderer-GaEB} are a useful tool to perform distributed optimization \cite{Li-CDC-ECC}, \cite{Zhang-book} thanks to their intrinsic properties: convergence to a pure Nash Equilibrium (NE) is always guaranteed and these equilibria are global maximizers of the potential function defined for the game. Therefore, if the potential function is defined as the network utility that we aim to maximize, we can optimize that network utility in a distributed way and achieve a solution closed to the optimal one that would be obtained solving exactly the MINLP problem. 

The main drawback of potential games when they are applied in a completely-distributed wireless scenario is that players usually require overall information about the remaining players of the network, making the solution not scalable. However, this limitation can be overcome with the cloud-based approach proposed in this work, since all the required information can be stored centrally and the game can be played using the computational resources of the cloud. 

Additionally, our game theoretic approach has two advantages when compared with solving directly the MINLP problem: first, its computational load is much lower, allowing a fast adaptation to changes in the network and decreasing the energy consumption to execute the algorithm, and second, it fits naturally into the distributed paradigm of cloud computing, since the players can be grouped into different virtual machines. The main contributions of our work are as follows:

\begin{enumerate}

\item We formally characterize the MINLP problem that arises when we try to maximize the proportional fairness of users in a wireless network. The decision variables in this problem are the serving access node, the downlink transmission power and the allocated channel. We consider that each access node has a limited backhaul capacity and a common set of channels, which causes intercell interference.

\item We propose two potential games to approximate the solution of the previous MINLP problem. In both games the potential function is defined so that the utility function of the players is completely aligned with the objective function of the MINLP problem. Additionally, we analyze the computational complexity and the convergence properties of both games.

\item We evaluate the proposed games by simulation and compare their performance with the optimal solution of the MINLP problem. This optimal solution is obtained with the branch-and-bound algorithm \cite{Belotti09}. 
\end{enumerate}

The remaining of the paper is organized as follows: Section \ref{sec:related_work} presents the related work. Section \ref{sec:model} describes the system model and the definitions of capacity that will be used. In Section \ref{sec:game} some basic concepts of game theory are given and the proposed games are explained in detail. Section \ref{sec:MINLP} presents the formal characterization of the MINLP problem that we aim to solve with the proposed games. Section \ref{sec:results} shows the simulation framework and the obtained results. Finally, some conclusions are provided in Section \ref{sec:conclusions}.

\section{Related Work}
\label{sec:related_work}
In this section we provide an overview of the contributions more related to our work, in particular those focused on joint resource allocation and cell selection; user association with backhaul constraints; and game-theoretic algorithms for resource allocation problems in wireless networks. 


User association, channel allocation and power control are typically the most determining factors for system performance in multicell wireless networks and therefore, many works have tackled some of these problems with different performance goals. Joint power control and user association \cite{Chen-Globecom09} -\nocite{Chen-ICC10}\cite{Chitti-SPAWC} and joint channel allocation and user association \cite{Fooladivanda-PIMRC11}, \cite{Fallgren-ICC11} have been thoroughly studied beforehand. However, a joint optimization of the three factors is a much less common topic \cite{Torregoza-EURASIP}, \cite{Li-ToVT}. In \cite{Torregoza-EURASIP} a multiobjective optimization problem is formulated to maximize the aggregated throughput of femtocell networks. In this problem, power control, base station assignment, and channel allocation are considered decision variables of a MINLP problem that are jointly optimized using the branch-and-bound algorithm. In \cite{Li-ToVT} a joint cell selection and power and channel allocation is performed to optimize the max-min throughput of all the cells of a network. Since solving directly this problem is unaffordable, authors propose an alternating optimization-based algorithm which applies branch-and-bound and simulated annealing. In their scenario, access nodes use the same power to communicate with all its associated users and fairness is taken into account to perform load balancing between access points, but not to allocate individual resources to the users. On the contrary, in our work we maximize the proportional fairness of the network, introduce backhaul restrictions and allow the allocation of several channels to the same user. Although we also arrive to a MINLP problem, it cannot be solved with the branch-and-bound algorithm in real time due to its complexity. For this reason, we propose a potential game which aims at maximizing the same objective function. This game converges to NEs close to the optimal solution with a much lower computational complexity than the branch-and-bound algorithm.

The cell selection problem when resources are also constrained by the backhaul network is now beginning to receive attention as well. In \cite{Domenico-PIMRC}, the authors present a cell selection framework which models the relationships among cell load, resource management, backhaul capacity constraints and the overall network capacity. In order to perform the cell selection scheme that maximizes the network capacity, they propose a centralized heuristic algorithm of limited complexity. Power control and channel allocation are not included in the cell selection process.


Game theory has been widely used to analyze resource allocation and cell selection problems in wireless networks \cite{Gao-Secon11} -\nocite{Malanchini-ToMC}\cite{Lin-ToWC}. In \cite{Malanchini-ToMC} the problem of access point selection and resource allocation is solved with a multileader/multifollower two-stage game. In the first stage, the access points (that belong to different network operators) play by choosing their resource allocation (a channel in this case) to maximize the number of users they serve, while in the second stage the users play a congestion game to select its serving access point. In \cite{Lin-ToWC} two cell selection games (without considering resource allocation) are proposed to model the behavior of nonsubscriber users in femtocell networks. Additionally, the existence of pure strategy Nash equilibria for those games are proven under feasible utility functions. In \cite{Gao-Secon11} a distributed cell selection and resource allocation mechanism performed by the mobile stations is presented. The problem is formulated as a two-tier game, named as inter-cell game and intra-cell game respectively. In the inter-cell game mobile stations perform cell selection, whereas in the intra-cell game they choose the proper radio resources in the serving cell. The existence of Nash equilibria of both games is analyzed and distributed algorithms to obtain mixed-strategy Nash equilibria are proposed. Nevertheless, their quality with regard to the optimum solution is not studied and it is assumed that there is no inter-cell interference. In addition, the objective function of the users is to maximize their own capacity, which can have a negative impact on the global throughput or the fairness of the network.

In this work, we bring together the joint problem of power control, channel allocation and user association with the backhaul constraints aiming to perform distributed optimization in a cloud-based approach. To tackle this problem we propose a game theoretic framework based on the design of potential games, where decisions are distributedly taken thanks to the information gathered and sent by the users. To ensure fairness, the objective function of each player is the sum of the logarithms of all the users' capacities. We evaluate the computational complexity of the proposed games and also the quality of the obtained equilibria by comparing with the global optimum solution of the equivalent MINLP problem, which demonstrates the benefits of the designed games to implement distributed, less costly solutions.




\section{System Model}
\label{sec:model}

The system considered is a downlink wireless network formed by a set $M$ of access nodes and a set $N$ of users. Each user $i$ is under the coverage of several access nodes, but it can be served only by one of them at the same time. The system has a discrete set $R$ of orthogonal resources, such as time and frequency slots in an OFDM signal. Each element of $R$, that in the following we name as channel, represents the minimum amount of resources that can be allocated to a user. The set of resources available at each access node, $R_j$, is a subset of $R$ and may vary from one access node to another one. Additionally, several access nodes may share the same resources, therefore $R_j \cap R_k \neq \null \emptyset$ for some access nodes $j$ and $k$ \footnote{In the case that $R_j = R, \forall j \in M$, we have a network with Full Frequency Reuse.}. Each access node must distribute its resources $R_j$ among the users served by it. Therefore, when a user $i$ is served by an access node $j$, $j$ allocates a subset of $R_j$ to $i$. It must be noted that according to the SINR model described in Section \ref{sec:access_model}, the resources can be completely reused by different access nodes, which will cause interference and a degradation of the SINR experienced by the users served with the same resources.

Regarding the backhaul, we assume that the connection of each access node to the core network has a limited capacity (non-ideal backhaul). Additionally, near access nodes can share the same backhaul capacity, which varies according to the geographic location. In order to achieve a proportional fairness in the resource allocation, our objective is to maximize the sum of the logarithms of the capacities allocated to all the users. However, in the considered scenario all the users are not guaranteed to access the network, and therefore some of their capacities could be zero. To allow zero-rate allocations, we modify the objective function, defining it as the sum of the logarithms of the capacities plus one \cite{Brehmer-book}.

\subsection{Wireless Access Capacity: SINR model}
\label{sec:access_model}

Let $c^{(a)}_{i,j}$ be the wireless access capacity that user $i$ obtains when it is connected to access node $j$. This capacity is defined as the sum of the capacities $c^{(c)}_{i,j,r}$ achieved in each channel $r$ of the subset $R_{i,j} \subseteq R_j$ of orthogonal channels that access node $j$ allocates to user $i$:

\begin{equation}
c^{(a)}_{i,j} = \sum_{r \in R_{i,j}} c^{(c)}_{i,j,r}
\end{equation}

These capacities $c^{(c)}_{i,j,r}$ are calculated using the physical interference model as follows. Given the transmission of access node $j$ to user $i$ in channel $r \in R_j$, the channel gain from $j$ to $i$ is defined as $g_{i,j} = d_{i,j}^{-\gamma}$, being $d_{i,j}$ the distance from $j$ to $i$ and $\gamma$ the path loss index. The transmission power is discretized into $Q+1$ levels $q = \left\lbrace 0, 1, \ldots, Q \right\rbrace$, equispaced between $0$ and the maximum transmission power $P_{max}$, which is the same for each channel. Thus, the transmission power used to reach user $i$ at channel $r$ by access node $j$ is:

\begin{equation}
p_{j,r} = q_{j,r} \cdot \frac{P_{max}}{Q}
\end{equation}

\noindent where $q_{j,r}$ represents the level $q$ at which AP $j$ transmits in channel $r$.

Under the physical interference model, a transmission is successful if the SINR at the receiving user is higher than a certain threshold $\alpha$, i.e, if it is fulfilled:

\begin{equation}
s_{i,j,r} = \frac{p_{j,r} \cdot g_{i,j}}{P_N + \sum_{\forall k \neq j}{p_{k,r} \cdot g_{i,k}}} \geq \alpha
\end{equation}

\noindent where $s_{i,j,r}$ is the SINR experienced by user $i$ at channel $r$ when it is served by access node $j$, $P_N$ is the background noise power and the terms $p_{k,r} \cdot g_{i,k}$ correspond to the interference from the rest of access nodes using the same channel $r$. 

Capacity $c^{(c)}_{i,j,r}$ will depend on the value of $s_{i,j,r}$. This capacity is upper-bounded by the theoretical limit obtained with the Shannon theorem, which states that the maximum achievable capacity in an AWGN (Additive White Gaussian Noise) channel is:

\begin{equation}
c^{(c)}_{i,j,r} = w_r \log_2 \left(1 + s_{i,j,r} \right)
\label{eq:shannon}
\end{equation}

\noindent with $w_r$ the bandwidth of channel $r$.

However, in a real system, the transmission rates are typically discrete and depend on a predefined set of modulation and channel coding schemes allowed in the system. The used transmission scheme is selected so that a specific bit error rate is guaranteed for the actual $s_{i,j,r}$, so generally speaking, $c^{(c)}_{i,j,r} = w_r f\left(s_{i,j,r}\right)$. A simplified method of introducing this effect in the proposed model is to define a discrete set of allowed values of spectral efficiency, $\eta = c^{(c)}/w$, and to obtain the associated SINR thresholds with (\ref{eq:shannon}) to build a step-wise function.

\subsection{Backhaul Capacity}
\label{sec:backhaul_model}

To model the backhaul from the access nodes to the core network, we assume that the access nodes are grouped into clusters, with all the access nodes in the same cluster $z$ sharing a backhaul capacity $C_z$. Typically, the formation of the clusters will depend on the geographical location of the access nodes. 

Let $c_{i,j}$ be the actual capacity allocated to user $i$ when connected to access node $j$ (i. e., the capacity taking into account backhaul restrictions), $N_j$ the set of users connected to access node $j$ (i.e. those users with $c_{i,j}^{(a)} \neq 0$) and $M_z$ the set of access nodes belonging to cluster $z$. As defined above, using a logarithmic utility function that allows zero-rate allocations, the optimization problem to solve for each cluster is:

\begin{equation}
\begin{tabular}{l l}
$  \max$ & $\displaystyle  \sum_{j \in M_z} \sum_{i \in N_j} \ln \left(1+c_{i,j} \right)$ \\
s.t. & $c_{i,j} \leq c^{(a)}_{i,j} \quad \forall i \in N_j, \forall j \in M_z$ \\
& $\displaystyle \sum_{j \in M_z} \sum_{i \in N_j} c_{i,j} \leq C_z$ \\
\end{tabular}
\label{eq:cluster_optimization}
\end{equation}

For simplicity, we assume that the data transmitted in the backhaul is the same as the data transmitted in the wireless access. Nevertheless, in a real system the data may differ depending on the processing performed in the access nodes (for instance, the data transmitted in the backhaul may be baseband signal or directly user data). To introduce this effect in our model, we should modify the first set of restrictions by multiplying the term $c_{i,j}^{(a)}$ with a constant $\beta$ which will depend on the specific processing performed in the access nodes.

The problem in (\ref{eq:cluster_optimization}) is equivalent to a max-min fairness problem \cite{Kelly-ETT} that can be trivially solved with the algorithm proposed in \cite{Bertsekas-book}. For the special case of the considered problem, a recursive expression for $c_{i,j}$ can be obtained. Let $\mathbf{c}^{(a)}_z$ be a vector containing the access capacities of the users connected to access nodes belonging to cluster $z$ sorted in non-decreasing order and $\mathbf{c}_z$ the vector of the corresponding actual capacities of these users. The actual capacity allocated to the user in the $k$-th position of these vectors is: 

\begin{equation}
c_z\left(k\right) = \begin{cases} c^{(a)}_z\left(k\right) & \mbox{if } c^{(a)}\left(k\right) \leq \displaystyle \frac{C_z - \sum _{ l < k} c_z\left(l\right)}{n_z - k + 1 } \\ \displaystyle \frac{C_z - \sum _{ l < k} c_z\left(l\right)}{n_z - k + 1 } & \mbox{otherwise } \end{cases} 
\end{equation} 

\noindent where $n_z = \sum _{ j \in M_z} \left| N_j \right|$ is the number of users connected to access nodes belonging to cluster $z$. Therefore, if the aggregated wireless access capacity of users in in the cluster is lower than $C_z$, then the actual capacity obtained by these users is the same as their wireless access capacity, that is, $c_{i,j} = c^{(a)}_{i,j}$. On the contrary, if this condition is not fulfilled, the backhaul capacity of the cluster must be shared between the users so that the log-sum of the their capacities is maximized. 

With these definitions, the design goal is to maximize the network utility (\textit{NU}), defined as the log-sum of the capacities $c_{i,j}$ plus one of all the users in the network in order to achieve optimal proportional fairness \cite{Brehmer-book}: 

\begin{equation}
NU = \sum_{i\in N} \ln\left(1+\sum_{j \in M} c_{i,j}\right)
\label{eq:objective}
\end{equation}

It is worth noting that $c_{i,j}$ can be higher than 0 only for one $j$ to ensure that each user is served only by one access node. That is, if $c_{i,j} \neq 0$ for some $j$, then $c_{i,k} = 0, \forall k \neq j$. This fact makes (\ref{eq:objective}) equivalent to:

\begin{equation}
NU = \sum_{i\in N} \sum_{j \in M} \ln\left(1+ c_{i,j}\right)
\label{eq:objective_bis}
\end{equation}

\section{Game Theoretic Solution}
 \label{sec:game}

As stated in Section \ref{sec:intro}, game theory can be used to achieve a good approximate solution for an optimization problem in a distributed manner. For this reason, we model the cell association and resource allocation problem as a formal game and perform the algorithmic design by correctly defining the set of players, the strategy space and the utility function.
 
Let be the game $\Gamma = \left\lbrace P,{\left\lbrace S_i \right\rbrace}_{i \in P}, {\left\lbrace u_i \right\rbrace}_{i \in P} \right\rbrace$, where $P$ is the finite set of players, $S_i$ is the set of strategies of player $i$ and $u_i : S \rightarrow \mathbb{R}$ is the utility function of that player, with $S = { \times _{i \in P}}{S_i}$ the strategy space of the game, formed by the Cartesian product of the set of strategies of each player in the game.

This utility function $u_i$ is a function of $s_i$, the strategy selected by player $i$, and of $s_{-i}$, the strategy profile of the rest of the players of the game. Each player will selfishly choose the strategy that improves its utility function considering the current strategies of the rest of players. 

One general key issue when designing a game is the choice of $u_i$ so that the individual actions of the players provide a good overall performance. In addition, in our specific scenario it is interesting the existence of an equilibrium point to ensure the convergence of the proposed algorithms when performing the optimization. In this context, it is useful the concept of Nash Equilibrium (NE), defined as a situation where no player has anything to gain by unilaterally deviating. Thus, a NE of a game $\Gamma$ is a profile ${s^*} \in S$ of actions such that for every player $i \in P$ we have:

\begin{equation}
	{u_i}(s_i^*,s_{ - i}^*) \ge {u_i}(s_i^{},s_{ - i}^*)
\end{equation}

\noindent for all $s_i \in S$, where $s_i$ denotes any strategy of player $i$ different from $s_i^*$ and $s_{-i}^*$ denotes the strategies of all the players other than player $i$ in the profile $s^*$. In our case, the convergence to a NE of the game makes it possible to reach a stable solution. In addition, the network can react to variations in the environment as any deviation from this equilibrium forces to play the game again to obtain a new NE.

\subsection{Players and Strategies}

To design our solution, the first decision that must be made is the definition of the players of the game (and their associated strategies). In this regard, we consider two different choices:

\begin{itemize}
\item The players are the $\left| N \right|$ users of the network ($P = N$). In this case, each strategy is the selection of an access node $j$ and the allocation of transmission power in each channel $r$ of $R$: $s_i = \left(j, p_{j,1}, p_{j,2}, \ldots, p_{j, \left| R \right|}\right)$. It must be noted that $p_{j,r} = 0 \; \forall r \notin R_j$. In the following we will denote this game as U-Game (\emph{User Game}).

\item Each player is an element of the set $P = \left\lbrace \left(x,y,z\right) | \; x \in N, y \in M_x, z \in R_y \right\rbrace$, with $M_x$ the set of access nodes that can serve user $x$ and $R_y$ the set of available channels for access node $y$. In this case, each strategy is the selection of the transmission power that will be used to serve user $x$ from access node $y$ in the channel $z$: $s_i = p_j$. In the sequel we will denote this game as C-Game (\emph{Channel Game}).
\end{itemize}

The main drawback of the U-Game is its complexity in terms of the computational load required to perform the strategy selection. As it will be seen in Section \ref{sec:complexity}, the complexity to select the strategy profile in the U-Game can be unaffordable.

It is worth noting that in both cases the actual players are not the physical users themselves. That is, physical users do not take the decision on the access node that serves them or the radio resources that are used to send them data. This decision is made by the network itself, that plays the game internally. The only task performed by the physical users is gathering and sending to the network the information required to estimate the channel gains between them and the access nodes close to them.  

\subsection{Utility Function and Convergence: Potential Game}
\label{sec:Utility}

An exact potential game is a game for which there exists a potential function $V : S \rightarrow \mathbb{R}$ such that:

\begin{equation}
\begin{split}
\Delta u_i & = u_i\left( s_i,s_{-i} \right) - u_i\left( s'_i,s_{-i} \right) = \Delta V = \\
& =  V \left( s_i,s_{-i} \right) - V \left( s'_i,s_{-i} \right) \; \; \forall i \in P,\forall {s_i},s{'_i} \in {S_i}
\end{split}
\end{equation}

This definition implies that each player's individual interest is aligned with the groups' interest (the potential function) since every change in the utility function of each player is directly reflected in the same change for the potential function. Therefore, any player choosing a better strategy given all other players’ current strategies will necessarily lead to an improvement in the value of the potential function. Thus, if only one player acts at each time step (repeated sequential game) and that player maximizes (best response strategy) or at least improves (better response strategy) its utility given the most recent action of the other players, then the process will always converge in finite steps to a NE \cite{Wang-ComNet}. In addition, global maximizers of the potential function $V$ are NE, although they may be just a subset of all NE of the game. 

These interesting properties of potential games (assured convergence in finite steps which can maximize the potential function) make them as useful tool to perform distributed optimization \cite{Li-CDC-ECC}, \cite{Zhang-book}. An important limitation to model a resource allocation problem as a potential game in a distributed wireless scenario is that players may require overall information about the remaining players of the network, making the solution not scalable \cite{Gallego-ADHOC}. However, as stated above, in our proposed framework the decisions are taken by the network itself, which collects the required information from the users. Therefore, we specifically design the game to be potential game.

We define a potential function for the considered scenario making $V$ the objective to maximize, in this case the network utility \textit{NU}. As for the utility function $u_i$, a direct option is to define it equal to the potential function (identical interest games \cite{Song-JSAC}):

\begin{equation}
u_i\left(s_i,s_{-i}\right) = \displaystyle\sum_{k \in N} \ln\left(1+\sum_{j \in M} c_{k,j}\right)
\label{eq:utility_1}
\end{equation}

This utility function is used for the U-Game and also for the C-Game with a slight modification explained in Section \ref{sec:Implementation}. In both cases the players need global information about all the access nodes and users in the network: to compute the utility function for each strategy, every player requires the channel gains $g_{l,m}$ between any user and their surrounding APs, the current selected strategies of the remaining players, and the available resources in the backhaul. Therefore, the actual implementation of the game must be performed in the network, where global information is available. 

\subsection{Implementation: Timing and Decision Rules}
\label{sec:Implementation}

A repeated sequential game with a round robin scheduling and a better response strategy is considered for the proposed games, which are played until a pure NE is found. Players evaluate first strategies with the lowest power profiles to reduce the interference over the remaining players. 

For the C-Game, the better response strategy must be modified to deal with the restriction that a user can be only served by one access node. If the round robin scheme is applied directly with this restriction, the strategy space of a player $i = \left(x_i,y_i,z_i\right)$ belonging to a user $x_i$ that is already served by a different access node $y_j \neq y_i$ should be $s_i = 0$. This situation can potentially lead to low quality NEs due to the impossibility of changing the serving access node of a user. To overcome this limitation, we divide the strategy selection process in the following two steps:

\begin{enumerate}
\item First, all the players belonging to the same user and access node (that is, with the same value of $x_i$ and $y_i$) play in order using a better response strategy. Additionally, they compute their utility function assuming that the user $x_i$ is only served by access node $y_i$ and that no other access node is transmitting to that user (that is, assuming that the interference generated in the network by players with $x_j = x_i$ and $y_j \neq y_i$ is zero):

\begin{equation}
u_i\left(s_i,s_{-i}\right) = \displaystyle\sum_{ \substack{ k \in N \\ k \neq x_i }} \ln\left(1+\sum_{j \in M} c_{k,j}\right) + \ln \left( 1+ c_{x_i, y_i}  \right)
\end{equation}

\item Once all the players corresponding to the same user and access node have selected their strategy (players of the form $\left(x_i, y_i, \cdot\right)$), we compare the network utility that is obtained when these players are transmitting to the network utility that is achieved when the players corresponding to the same user and a different access node are transmitting (players of the form $\left(x_i, y_j, \cdot\right)$). If the network utility has been improved, then we change the strategy of the players $\left(x_i, y_j, \cdot\right)$ to not transmit and leave the players $\left(x_i, y_i, \cdot\right)$ with their selected strategies. If not, we do the opposite. 

\end{enumerate}





This modification still ensures convergence to a NE: In a potential game, an improvement in the utility function of a player implies directly the same improvement in the potential function. With this strategy selection process, we ensure also that if the access node that serves a user is changed, then the potential function is also increased.

\subsection{Complexity of the Games}
\label{sec:complexity}

With the better response strategy described previously, each player tries to improve its current utility at each step regardless of the past history. Therefore, the complexity of the games is directly related to four factors:

\begin{itemize}
\item \emph{Number of rounds required to reach a stable point.}
\item \emph{Number of steps at each round}: Since a round robin strategy is followed, this number is equal to the number of players. For the U-Game, this value is $\left| N \right|$, while for the C-Game it is  $\sum_{i \in N} \sum_{j \in M_i} \left| R_j \right|$. This value can be upper bounded by $ m_{M} \left| N \right| \left| R \right|$, with $m_M$ the maximum number of access nodes that can cover a user. 
\item \emph{Number of possible strategies at each step}: For the U-Game, each step of the game may require in the worst case exploring $ \sum_{j \in M_i} Q^{\left| R_j \right|} \leq m_M Q ^{\left| R \right|}$ different strategies, which correspond to all the possible combinations of the $Q$ power levels at each of the $\left| R \right|$ channels available in the $m_M$ access nodes that may serve the user. For the C-Channel, this figure is reduced to $Q$ strategies, corresponding to all the power levels at each channel.
\item \emph{Computational complexity of calculating the utility function of each strategy}: We use as a reference to compare both games the number of channel capacities (terms $c^{(c)}_{i,j,r}$) required to calculate the utility function. This channel capacity is directly related to the SINR in the channel, as shown in Section \ref{sec:access_model}. For the U-Game, the evaluation of each strategy requires at most the calculation of $\left| N \right| \left| R \right|$ terms $c^{(c)}_{i,j,r}$, corresponding to all the channel capacities of all the users in the network. This value is the same for the C-Game.
\end{itemize}

Taken into account the last three factors, an upper bound for the computational complexity per round is given by:

\begin{itemize}
\item \emph{U-Game}: $ m_{M}  \left| N \right|^2 \left| R \right| Q ^{\left| R \right|}  $
\item \emph{C-Game}: $ m_{M}  \left| N \right|^2  \left| R \right|^2 Q $
\end{itemize}

Therefore, the computational load per round is much lower in the C-Game than in the U-Game since there is no exponential dependence on $\left | R \right|$. Regarding the first factor (the number of rounds required to reach an stable point), it will be analyzed in Section \ref{sec:results}. Note that for both cases it is guaranteed that the game will converge in a finite number of steps to a NE, as explained in section~\ref{sec:Utility}. With those results and the analysis performed in this section the computational complexity of the two games will be completely compared.

\section{Optimal Solution}
 \label{sec:MINLP}

In this Section we formulate the MINLP problem that must be solved to obtain the maximum value of the network utility defined in (\ref{eq:objective}). This value will be used to evaluate the quality of the proposed games. To formulate the MINLP problem, we proceed as follows: 

Let $M_i$ be the subset of $M$ containing the access nodes that can serve user $i$. For each user $i$, we define $\left|M_i\right|$ binary variables, $x_{i,j}$, indicating whether the access node $j$ is transmitting to user $i$ or not. Since each user can be served only by one access node, the following restrictions must be satisfied:

\begin{equation}
	 	\sum_{j \in M_i} x_{i,j} \leq 1 \qquad \forall i \in N
	 	\label{Eq_Op_1}
\end{equation} 

For each pair $(i,j)$ we also define $\left|R_j\right|$ binary variables, $y_{i,j,r}$, indicating whether access node $j$ is transmitting to user $i$ in channel $r$ or not. Since an access node can allocate several channels to the same user and each user can be served only by one access node, $y_{i,j,r}$ can only be 1 if the corresponding $x_{i,j}$ is also 1. This can be expressed as follows: 

\begin{equation}
	 	\ y_{i,j,r} \leq x_{i,j} \qquad \forall i \in N, \forall j \in M_i, \forall r \in R_j
	 	\label{Eq_Op_2}
\end{equation} 

Additionally, we assume that each channel available at an access node $j$ cannot be shared between two different users, which implies that:

\begin{equation}
	 	\sum_{i \in N} y_{i,j,r} \leq 1 \qquad \forall j \in M, \forall r \in R_j
	 	\label{Eq_Op_1b}
\end{equation}

As stated in Section \ref{sec:model}, the transmission power is discretized into $Q+1$ levels $q = \left\{0, 1, \dotsc , Q\right\}$. Let us define the variables $q_{i,j,r}$ indicating the level $q$ at which access node $j$ transmits to user $i$ in channel $r$. According to the definition of $y_{i,j,r}$, $q_{i,j,r}$ must be 0 if $y_{i,j,r} = 0$ and it must be comprised between 1 and $Q$ if $y_{i,j,r} = 1$. These restrictions can be expressed mathematically as:

\begin{equation}
	 	y_{i,j,r} \leq q_{i,j,r} \leq Q \cdot y_{i,j,r} \qquad \forall i \in N, \forall j \in M_i, \forall r \in R_j
	 	\label{Eq_Op_3}
\end{equation} 

We also define the variables $s_{i,j,r}$, which correspond to the SINR at user $i$ when it is served by access node $j$ at channel $r$. These variables are function of the transmission powers of the access node $j$ at channel $r$ and the rest of interfering access nodes using the same channel:

\begin{equation}
s_{i,j,r} = \frac{q_{i,j,r}\cdot \displaystyle\frac{P_{max}}{Q} \cdot g_{i,j}}{P_N+\displaystyle\sum_{\substack{l \in M \\ l \neq j} }\sum_{k \in N }{ q_{k,l,r}\cdot \frac{P_{max}}{Q} \cdot g_{i,l}}} \quad \forall i \in N, \forall j \in M_i, \forall r \in R_j
\label{Eq_Op_4}
\end{equation}


This expression can be manipulated to transform the division into multiplications, which are more adequate for mathematical programming: 

\begin{equation}
\begin{tabular}{c}
$P_N \cdot s_{i,j,r} + \displaystyle \sum_{\substack{l \in M \\ l \neq j} } \sum_{k \in N } {g'_{i,l} \cdot q_{k,l,r} \cdot s_{i,j,r}} -  g'_{i,j} \cdot q_{i,j,r} = 0$ \\
\multicolumn{1}{r}{$\forall i \in N, \forall j \in M_i, \forall r \in R_j$}
\end{tabular} 
\label{Eq_Op_5}
\end{equation}

\noindent with $g'_{i,l} = g_{i,l} \! \cdot \! P_{max} / Q$. To solve the MINLP problem that we are formulating with the branch-and-bound algorithm, we must apply a Reformulation-Linerization Technique (RLT) to the non-linear cross-products $q_{k,l,r} \cdot s_{i,j,r}$. This technique is required to obtain a convex hull representation of the non-linear terms. To do so, we define first the variables $t_{l,r}$ denoting the total power transmitted by access node $l$ in channel $r$:
\begin{equation}
t_{l,r}=\displaystyle \sum_{k \in N} q_{k,l,r}
\end{equation}

With these variables, we can rewrite (\ref{Eq_Op_5}) as:
\begin{equation}
\begin{tabular}{c}
$P_N \cdot s_{i,j,r} + \displaystyle \sum_{\substack{l \in M \\ l \neq j} } {g'_{i,l} \cdot t_{l,r} \cdot s_{i,j,r}} -  g'_{i,j} \cdot q_{i,j,r} = 0$ \\
\multicolumn{1}{r}{$\forall i \in N, \forall j \in M_i, \forall r \in R_j$}
\end{tabular} 
\label{Eq_Op_5b}
\end{equation}

Now we introduce the auxiliary variables $u_{i,j,l,r} = t_{l,r} \cdot s_{i,j,r}$ and apply the well-known linearization inequalities proposed by McCormick \cite{McCormick76}:

\begin{equation}
\begin{tabular}{c}
$u_{i,j,l,r} \geq t_{l,r}^L \cdot s_{i,j,r} +  s_{i,j,r}^L \cdot t_{l,r} - t_{l,r}^L \cdot s_{i,j,r}^L$\\ 
$u_{i,j,l,r} \geq t_{l,r}^U \cdot s_{i,j,r} +  s_{i,j,r}^U \cdot t_{l,r} - t_{l,r}^U \cdot s_{i,j,r}^U$\\ 
$u_{i,j,l,r} \leq t_{l,r}^L \cdot s_{i,j,r} +  s_{i,j,r}^U \cdot t_{l,r} - t_{l,r}^L \cdot s_{i,j,r}^U$\\ 
$u_{i,j,l,r} \leq t_{l,r}^U \cdot s_{i,j,r} +  s_{i,j,r}^L \cdot t_{l,r} - t_{l,r}^U \cdot s_{i,j,r}^L$
\label{Eq_Op_6}
\end{tabular} 
\end{equation}

\noindent where the superscript $U$ or $L$ represents the upper or lower bound of a variable. It must be noted that these restrictions will be updated in each step of the branch-and-bound algorithm since the bounds of the variables $q_{k,l,r}$ (and therefore of the variables $t_{l,r}$ and $s_{i,j,r}$ which depend of them) can vary in each step of the algorithm.

Regarding the SINR restriction, we must ensure that if access node $j$ is transmitting to user $i$ in channel $r$ ($y_{i,j,r}~=~1)$, then the SINR for this channel, $s_{i,j,r}$, must be higher than $\alpha$. On the contrary, if $s_{i,j,r}$ is lower than $\alpha$, we must set $y_{i,j,r}$ and $q_{i,j,r}$ to 0 to decrease the interference of the system. These conditions can be expressed with the following inequalities:  
 
\begin{equation}
	 	s_{i,j,r} \geq \alpha \cdot y_{i,j,r} \qquad \forall i \in N, \forall j \in M_i, \forall r \in R_j
	 	\label{Eq_Op_7}
\end{equation} 

Related to the variables $s_{i,j,r}$, we also define the variables $c^{(c)}_{i,j,r}$ that indicate the capacity obtained by user $i$ in channel $r$ when it is served by access node $j$. As stated in Section~\ref{sec:model}, we assume that these capacities can only take a set of discrete values dependent of some SINR thresholds. This makes the relationship between $c^{(c)}_{i,j,r}$ and $s_{i,j,r}$ be given by a stepwise function of the form:

\begin{equation}
	 	c^{(c)}_{i,j,r} = f\left( s_{i,j,r} \right) = \left\{   \begin{array}{l l} 0 & \quad \text{if $s_{i,j,r} \leq s_{th_1} $} \\
    c_1 & \quad \text{if $s_{th_1} < s_{i,j,r} \leq s_{th_2}$} \\  c_2 & \quad \text{if $s_{th_2} < s_{i,j,r} \leq s_{th_3}$} \\ & \vdots \\ c_C & \quad \text{if $ s_{th_C} < s_{i,j,r}$} \\ \end{array} \right.
	 	\label{Eq_Op_8}
\end{equation} 

\noindent where $C$ is the number of available capacities and $s_{th_1} =\alpha$. Again, a linearization of this function is needed to solve the MINLP problem. To do so, we propose two different sets of inequalities: first, a lower and an upper bound for $c^{(c)}_{i,j,r}$ imposed by the lower and upper bounds of $s_{i,j,r}$: 

\begin{equation}
\begin{array}{c}
c^{(c)}_{i,j,r} \geq f\left(s_{i,j,r}^L \right) \quad \forall i \in N, \forall j \in M_i, \forall r \in R_j \\ 

c^{(c)}_{i,j,r} \leq f\left(s_{i,j,r}^U \right) \quad \forall i \in N, \forall j \in M_i, \forall r \in R_j
\end{array} 
\label{Eq_Op_11}
\end{equation}

These bounds for $c^{(c)}_{i,j,r}$ will be narrowed as $s_{i,j,r}^L$ and $s_{i,j,r}^U$ get closer during the execution of the branch-and-bound algorithm. The second set of inequalities is based on the lines formed by two consecutive points $\left\{ \left( s_{th_m}, c_m\right), \left(s_{th_{m+1}}, c_{m+1}\right) \right\}$ of function $f$:

\begin{equation}
c^{(c)}_{i,j,r} \leq a_m \cdot s_{i,j,r} + b_m \qquad \forall i \in N, \forall j \in M_i, \forall r \in R_j, 0\leq m < C 
\label{Eq_Op_12}
\end{equation}

\noindent with

\begin{equation}
 a_m =  \displaystyle \frac{c_{m+1} - c_m}{s_{th_{m+1}} - s_{th_m}}
\label{Eq_Op_13}
\end{equation}

\begin{equation}
 b_m =  \displaystyle \frac{ s_{th_{m+1}} \cdot c_m - s_{th_m} \cdot c_{m+1} } {s_{th_{m+1}} - s_{th_m}}
\label{Eq_Op_14}
\end{equation}

\noindent For the first line passing through the origin, we consider $c_0 = 0$ and $s_{th_0} = 0$. 

Finally, we define the variables $c_{i,j}$ representing the capacity obtained by user $i$ when connected to access node $j$. This capacity cannot be higher than the total capacity achievable in the channels allocated to $i$ by $j$:

\begin{equation}
c_{i,j} \leq \sum_{r \in R_j} c^{(c)}_{i,j,r} \qquad \forall i \in N, \forall j \in M_i
\label{Eq_Op_10}
\end{equation}

Additionally, $c_{i,j}$ is also bounded by the total capacity available in the backhaul, that must be shared between the users connected to the same cluster $z$:

\begin{equation}
\sum_{i\in N_z} c_{i,j} \leq c^{(b)}_z  \qquad \forall z \in Z
\label{Eq_Op_9}
\end{equation}

\noindent where $c^{(b)}_z$ is the backhaul capacity of zone $z$ and $Z$ is the set of backhaul zones. According to the logarithmic utility definition defined in section~\ref{sec:model}, the objective function can be expressed in terms of $c_{i,j}$ as:

\begin{equation}
\sum_{i\in N} \ln\left(1+\sum_{j \in M} c_{i,j}\right)
\label{Eq_Op_15}
\end{equation}

To obtain a linearization of the objective function we use the following linear relaxation for the logarithmic function \cite{Shi11}. Let us assume that the upper and lower bounds of $c_i = \sum_{j \in M} c_{i,j}$ are $c_{i}^U$ and $c_{i}^L$ respectively. Then the function $\ln(1+c_{i})$ lies inside the convex hull defined by the chord of the function between the points $\left(c_{i}^L, \ln\left(1+c_{i}^L\right)\right)$ and $\left(c_{i}^U, \ln\left(1+c_{i}^U\right)\right)$, and three segments tangential to the logarithm at the points $\left(c_{i}^L, \ln\left(1+c_{i}^L\right)\right)$, $\left(c_{i}^U, \ln\left(1+c_{i}^U\right)\right)$ and $\left(\beta,\ln\left(1+\beta\right)\right)$ with $\beta$ the value of $c_i$ at the intersection point of the other two segments:

\begin{equation}
\beta = \displaystyle \frac{\left[1+c_{i}^L\right]\cdot\left[1+c_{i}^U\right]\cdot\left[\ln(1+c_{i}^U)-\ln(1+c_{i}^L)\right]}{c_{i}^U-c_{i}^L}-1
\label{Eq_Op_16}
\end{equation}

Therefore, the convex region can be described by the following four linear constraints:

\begin{equation}
\begin{array}{c}

\left[c_{i}^U-c_{i}^L\right] \cdot \ln\left(1+c_{i}\right) + \left[\ln\left(1+c_{i}^L\right)-\ln\left(1+c_{i}^U\right)\right]\cdot c_{i} \geq c_{i}^U\cdot \ln\left(1+c_{i}^L\right) -  c_{i}^L\cdot \ln\left(1+c_{i}^U\right)\\

\left[1+c_{i}^L\right]\cdot \ln\left(1+c_{i}\right)- c_{i} \leq \left[1+c_{i}^L\right] \cdot \ln\left(1+c_{i}^L\right) - c_i^L\\

\left[1+c_{i}^U\right] \cdot \ln\left(1+c_{i}\right)-c_{i} \leq \left[1+c_{i}^U \right]\cdot\ln \left(1+c_{i}^U\right) - c_i^U\\

\left[1+\beta\right] \cdot \ln\left(1+c_{i}\right) -c_{i} \leq \left[1+\beta\right]\cdot\ln\left(1+\beta\right) - \beta  \\

\end{array}
\label{Eq_Op_17}
\end{equation}

\begin{figure*}\footnotesize
\centering
\renewcommand{\arraystretch}{1.3}
\begin{tabular}{|l l l|}
\hline
  Maximize  & $\sum_{i\in N} \ln\left(1+\sum_{j \in M} c_{i,j}\right)$ & \\
  Constraints  & $\sum_{j \in M_i} x_{i,j} \leq 1 $  & $\forall i \in N$ \\
    & $\ y_{i,j,r} \leq x_{i,j} $ & $\forall i \in N, \forall j \in M_i, \forall r \in R_j$ \\
& 	 	$\sum_{i \in N} y_{i,j,r} \leq 1$  & $\forall j \in M, \forall r \in R_j    $ \\
    
    & $y_{i,j,r} \leq q_{i,j,r} \leq Q \cdot y_{i,j,r}$ & $\forall i \in N, \forall j \in M_i, \forall r \in R_j$ \\
    & $t_{l,r}=\sum_{k \in N} q_{k,l,r}$ & $\forall l\in M, \forall r \in R_l$\\ 
     & $P_N \cdot s_{i,j,r} + \sum_{\substack{l \in M \\ l \neq j} }{g'_{i,l} \cdot u_{i,j,l,r} } -  g'_{i,j} \cdot q_{i,j,r} = 0$ & $\forall i \in N, \forall j \in M_i, \forall r \in R_l$\\
    & Linear constraints for $u_{i,j,l,r}$ & $\forall i \in N, \forall j \in M_i, \forall l \in M, \forall r \in R_l$ \\
& $s_{i,j,r} \geq \alpha \cdot y_{i,j,r}  $ & $\forall i \in N, \forall j \in M_i, \forall r \in R_j$ \\
& Linear constraints for $c^{(c)}_{i,j,r}$ & $\forall i \in N, \forall j \in M_i, \forall r \in R_j$\\
& $c_{i,j} \leq \sum_{r \in R_j} c^{(c)}_{i,j,r} $ &  $\forall i \in N, \forall j \in M_i $ \\
& $\sum_{i\in N_z} c_{i,j} \leq c^{(b)}_z  $ & $\forall z \in Z$ \\
& Linear constraints for $\ln\left(1+\sum_{j \in M} c_{i,j}\right)$ & $\forall i \in N$ \\

\hline  
\end{tabular}
\caption{Problem formulation}
\label{Fig_problem}
\end{figure*}

Figure \ref{Fig_problem} summarizes the optimization problem, being the variables $x_{i,j}$ and $y_{i,j,r}$ binary, $q_{i,j,r}$ integer in the range $\left\{0, \dotsc, Q\right\}$ and $s_{i,j,r}$, $t_{l,r}$, $u_{i,j,l,r}$, $c^{(c)}_{i,j,r}$ and $c_{i,j}$ real numbers. This problem can be solved to obtain the global optimal solution applying the branch-and-bound algorithm \cite{Belotti09}, \cite{Ortin-ADHOC}, which relaxes the variables $x_{i,j}$, $y_{i,j,r}$ and $q_{i,j,r}$ and treats them as real variables. The relaxed problem is located at the root node of a tree that the branch-and-bound algorithm generates dynamically to solve the original problem. Each node of the tree will be composed of this initial relaxed problem with appended restrictions that generates partitions of its solution space.

\section{Results}
\label{sec:results}

The analysis and evaluation of a game model should cover two different aspects: first, the existence of some equilibrium points and second, their quality, which can be measured as the ratio between the network utility obtained in the equilibrium and the maximum achievable network utility. Concerning the first issue, the convergence to an equilibrium of the proposed games has been analyzed in Section \ref{sec:game}. Given the well-known properties of the potential games \cite{Monderer-GaEB}, the existence and convergence to a NE is guaranteed for both games. As for the quality of the equilibria, the proposed games have been evaluated by simulation and compared to the optimal solution that would be obtained solving the MINLP problem described in section \ref{sec:MINLP}\footnote{The problem has been solved using the CPLEX software \cite{cplex}}.

To analyze the proposed games, several scenarios have been studied varying the values of the main simulation parameters: topology size, number of users and access nodes, available frequency channels and power levels. The relative differences among the two games hold in any case. On the other hand, the computational complexity of solving the MINLP problem makes it difficult to obtain results as the network size grows. For this reason, we present results in two different scenarios: a first scenario to compare the performance of the two proposed games and a simplified second scenario to compare the outcome of the games with the optimum solution.

In the first scenario, the network consists of 4 equispaced access nodes in a square area of 200 $\times$ 200 $\text{m}^2$ in positions (50, 50), (100,50), (50, 150) and (150, 150). Several numbers of users (ranging from 4 to 20) are randomly deployed in that area. The total number of orthogonal channels $R$ is 8. The subset of channels allocated to each access node $j$, $R_j$, is a random subset between 3 and 7 of the channels in $R$. There are $B$ = 4 different geographic zones with different backhaul capacities. Each access node belongs to a different zone. The backhaul capacity of each zone is randomly selected among three possible values: 10, 20 and 30 Mbps. All the results for the games are averaged with 1000 random instances of the scenario. 

$P_{max}$ is set to 20 dBm, there are $Q$ = 4 different levels of transmission power and 7 values of spectral efficiency (1, 1.5, 2, 3, 4, 4.5 and 6). The bandwidth of each channel, $w_r$, is 1 MHz, the path loss index is $\gamma$ = 4.5 and the noise power $P_N$ is -105 dBm. The SINR threshold $\alpha$ is set to 0 dB, which corresponds to the SINR required to obtain the minimum spectral efficiency $\eta$ = 1. With these parameters and in the absence of interference, the maximum distance from one access node at which the highest spectral efficiency can be obtained is 238 meters, so any user could obtain it regardless of its position.

The following results analyze 1) the performance of the proposed games as efficient cell selection and channel and power allocation mechanisms and 2) the fairness of the considered utility function. To this second purpose, both games have been simulated with the utility function defined in (\ref{eq:utility_1}), shown with the legend \emph{log} in the graphs, and with a utility function focused only on maximizing the network capacity:

\begin{equation}
u_i\left(s_i,s_{-i}\right) = \displaystyle\sum_{k \in N} \left(\sum_{j \in M} c_{k,j}\right)
\label{eq:utility_3}
\end{equation}

\noindent shown with the legend \emph{cap}. This latter utility function, which also fulfils the potential condition, is defined here to compare the fairness of the proposed log-sum utility function with the fairness of a similar resource allocation scheme designed exclusively to maximize the total throughput of the network.

In addition, the performance when a user can be served by any of the four access nodes (4AN in the graphs) is compared to the case when it can be served only by the nearest one (1AN). In all the cases, a better response strategy until reaching a NE has been followed.

\begin{figure}[h]
\centering
\includegraphics[scale=0.5]{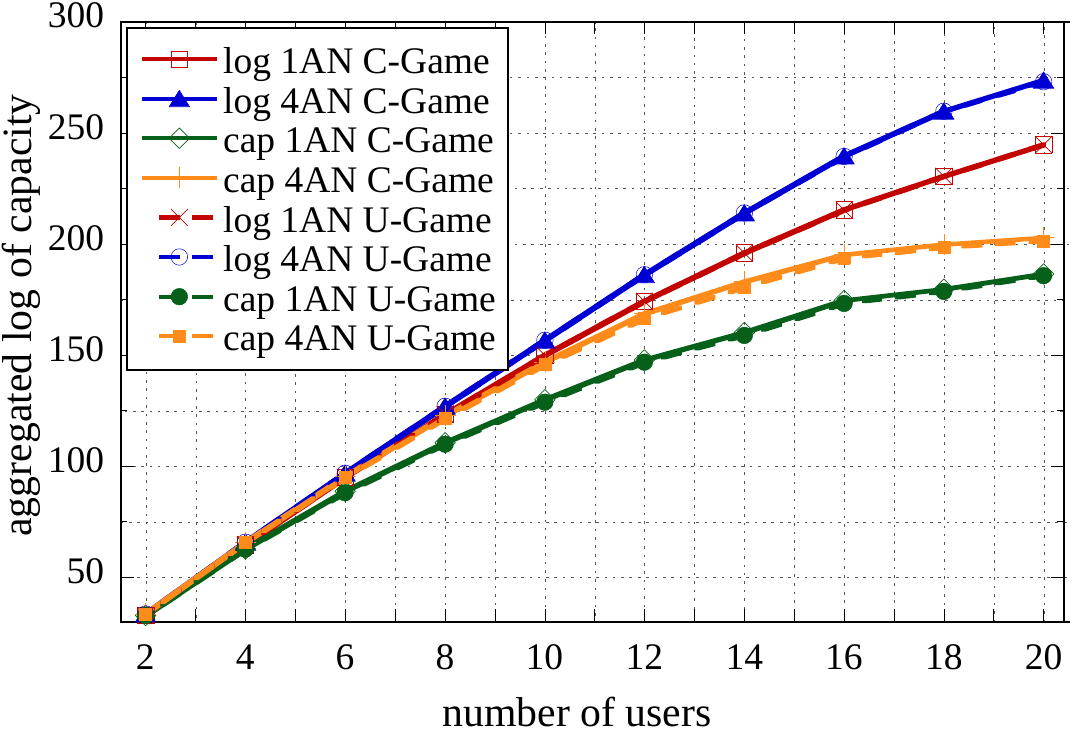}
\caption{Aggregated log-capacity for the C-Game and the U-Game with different utility functions and association policies.}
\label{fig:Log-games}
\end{figure}

\begin{figure}[h]
\centering
\includegraphics[scale=0.5]{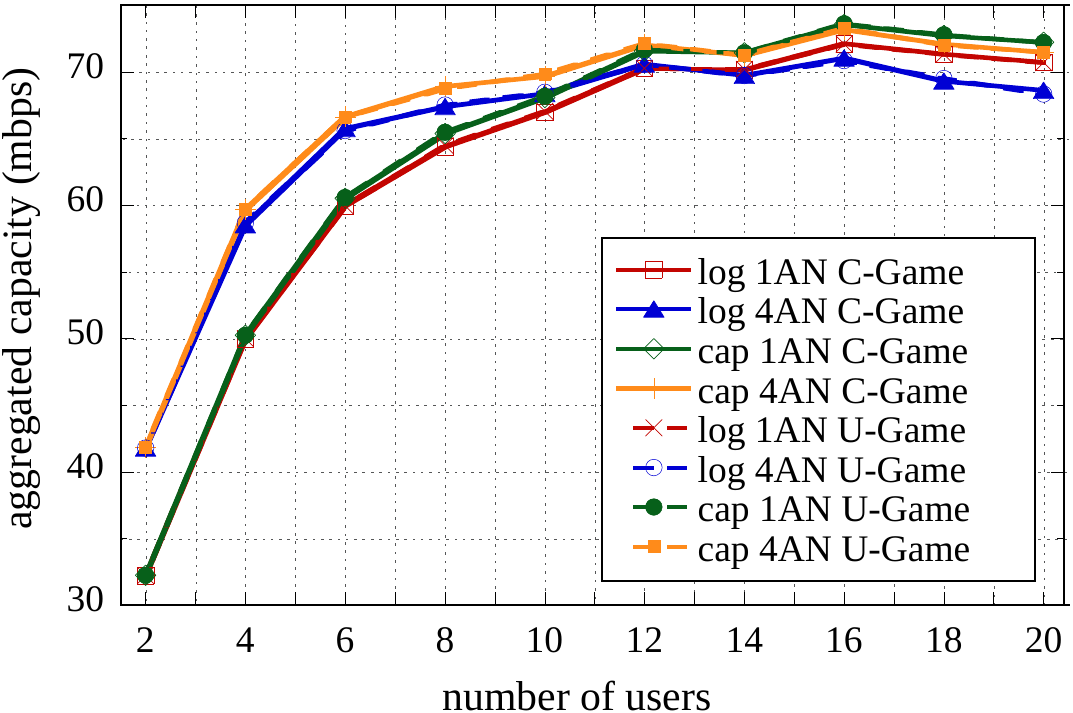}
\caption{Aggregated capacity for the C-Game and the U-Game with different utility functions and association policies.}
\label{fig:Capacity-games}
\end{figure}

Fig.~\ref{fig:Log-games} and Fig.~\ref{fig:Capacity-games} show the aggregated sum of the logarithms of the users' capacities (aggregated log-capacity hereafter) according to (\ref{eq:utility_1}) and the aggregated network capacity according to (\ref{eq:utility_3}). As can be seen, the C-Game provides basically the same performance as the more complex U-Game. Additionally, the fact of selecting the access node improves significantly the aggregated log-capacity, which will increase the fairness of the network as it will be shown shortly.

\begin{figure}[h]
\centering
\includegraphics[scale=0.5]{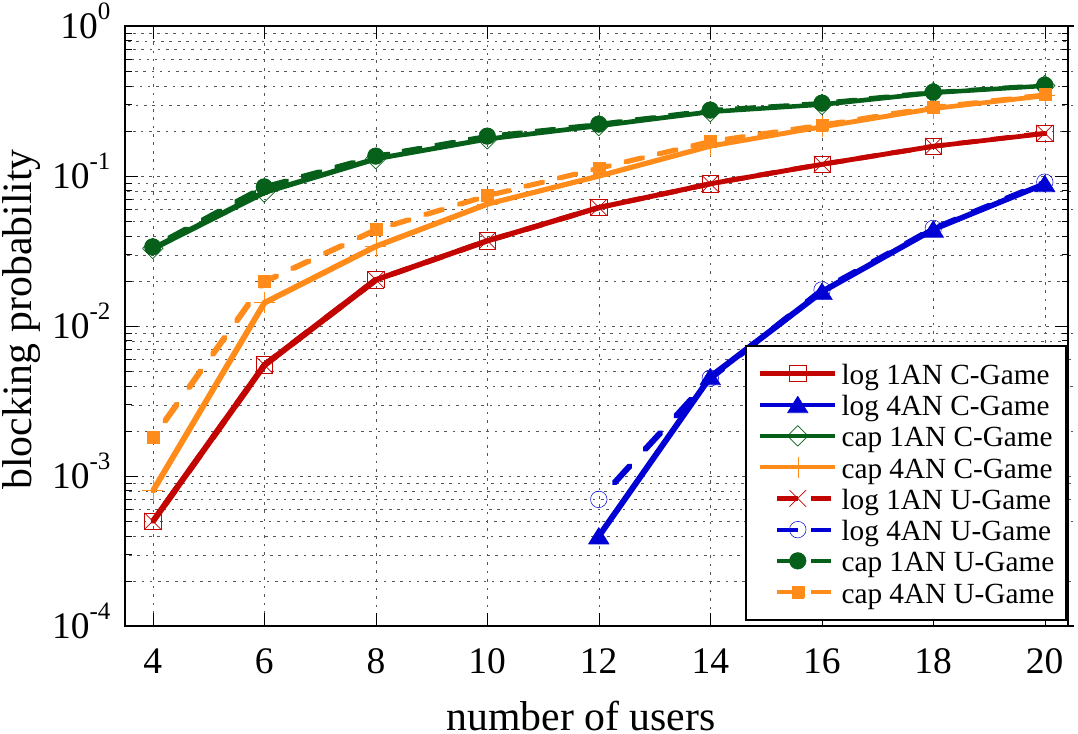}
\caption{Blocking probability for the C-Game and the U-Game with different utility functions and association policies.}
\label{fig:Block-games}
\end{figure}

\begin{figure}[h]
\centering
\includegraphics[scale=0.5]{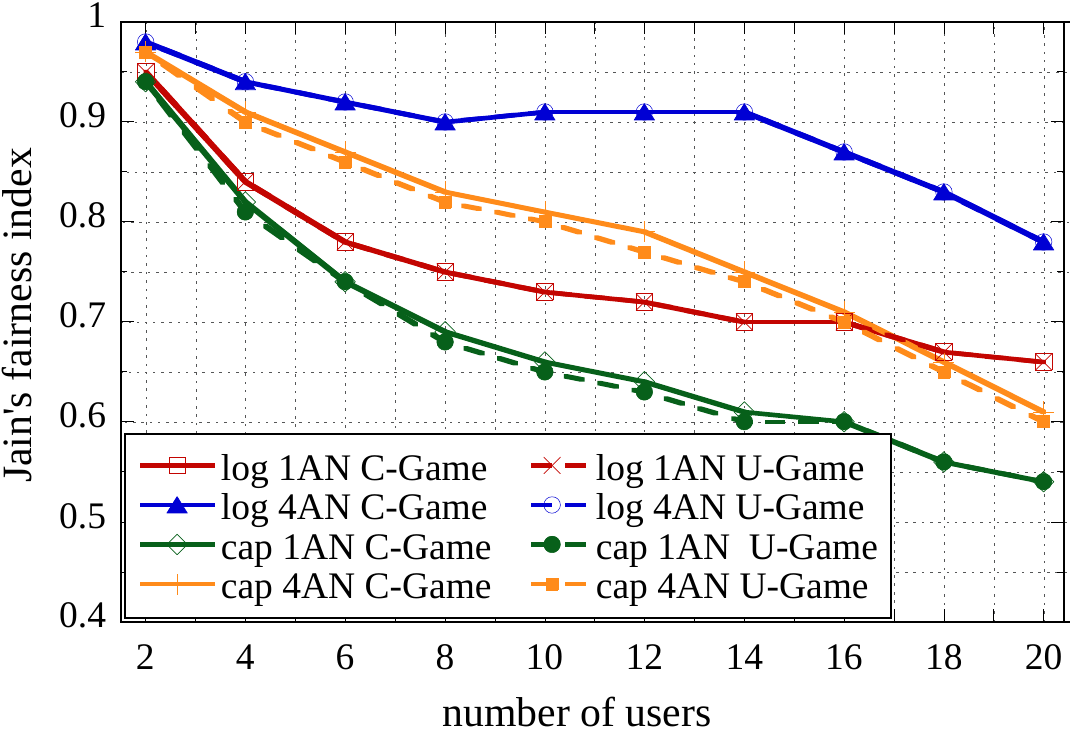}
\caption{Jain's fairness index for the C-Game and the U-Game with different utility functions and association policies.}
\label{fig:Jain-games}
\end{figure}

To measure the fairness of the proposed games, we show the blocking probability for users trying to access the network in Fig.~\ref{fig:Block-games} and the Jain's index in Fig.~\ref{fig:Jain-games}.
The Jain's index provides a quantitative measure of the fairness achieved in a network and is calculated as:

\begin{equation}
J = \frac{{{{\left( {\sum\limits_{i \in N} {{c_i}} } \right)}^2}}}{{\left| N \right|\sum\limits_{i \in N} {{c_i}^2} }}
\label{eq:jain}
\end{equation}

\noindent where $c_i$ denotes the capacity obtained by user $i$.

As shown in these figures, both the use of the logarithmic utility function and the capability of selecting the access node provide a marked improvement in the network fairness, reducing the blocking probability and achieving a fairer sharing of the available resources according to the Jain's index without excessively reducing the overall network capacity.

\begin{figure}[h]
\centering
\includegraphics[scale=0.5]{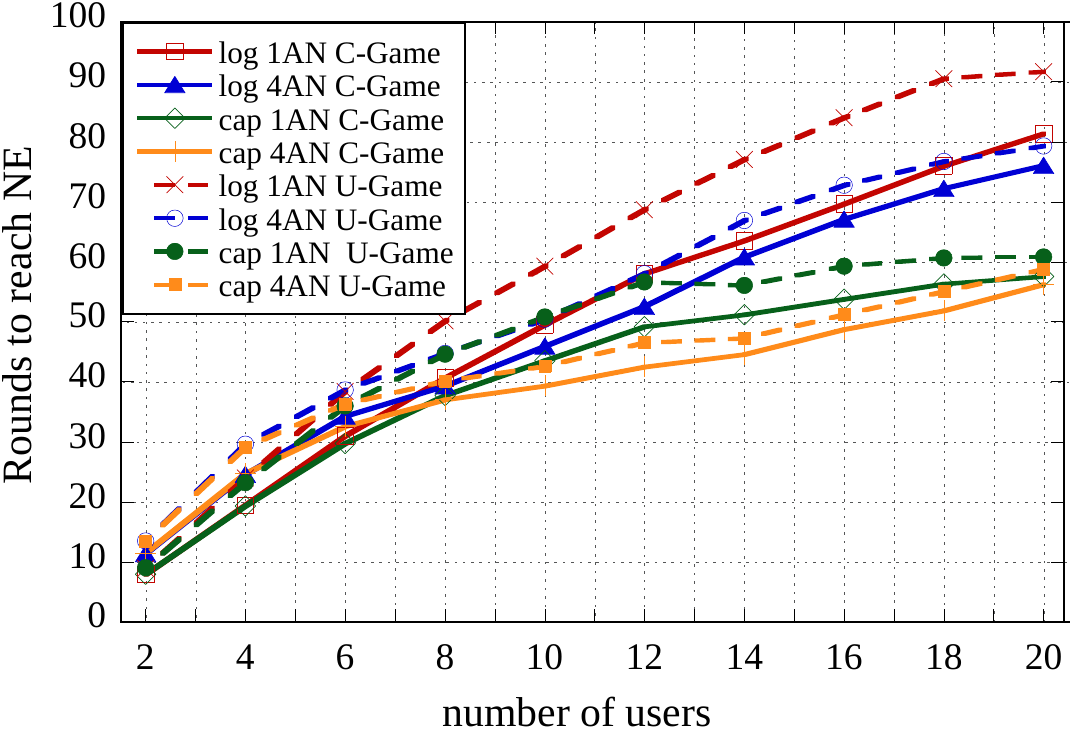}
\caption{Number of rounds required to reach NE for the C-Game and the U-Game with different utility functions and association policies.}
\label{fig:Rounds-games}
\end{figure}

Fig.~\ref{fig:Rounds-games} shows the mean number of rounds to reach a stable point for the games with the same parameters considered in the previous results. As can be seen, the mean number of rounds is very similar for both games, which confirms the overall lower complexity of the C-Game compared to the U-Game. For example, in the simulated scenario the upper bound for the computational complexity per round with $N = 20$ users and the possibility of performing the cell selection up to with four access nodes is $4 \cdot 20^2 \cdot 7 \cdot 4^7 = 1.835 \cdot 10^8$ for the U-Game and $4 \cdot 20^2 \cdot 7^2 \cdot 4 = 3.136 \cdot 10^5$ for the C-Game.

\begin{figure}[h]
\centering
\includegraphics[scale=0.5]{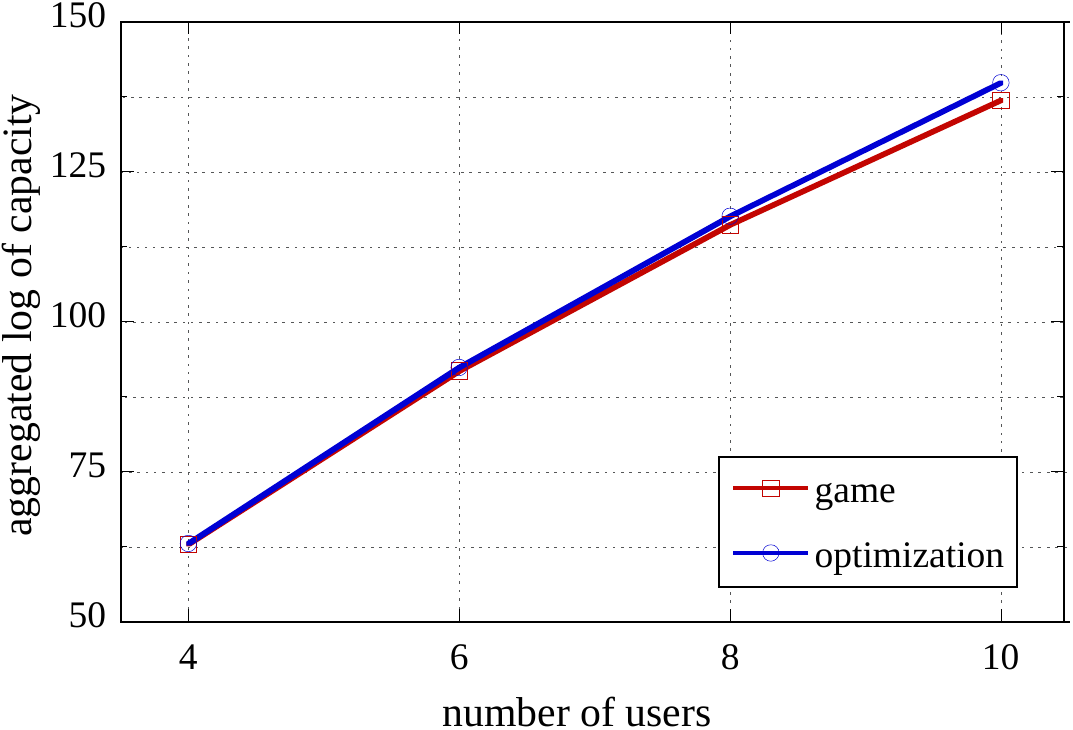}
\caption{Aggregated log-capacity for the C-Game vs. optimum solution.}
\label{fig:Log-opt}
\end{figure}

In order to compare the proposed games with the optimal solution and due to the complexity of solving the MINLP problem, in the second scenario the number of orthogonal channels $\left|R\right|$ is reduced to 3 (available to all the access nodes) and the value of $Q$ is set to 2. The optimal solution is compared to that of the C-Game considering the logarithmic utility function and that a user can be served by any of the four access nodes. All the results are averaged with 100 random instances of the scenario.

\begin{figure}[h]
\centering
\includegraphics[scale=0.5]{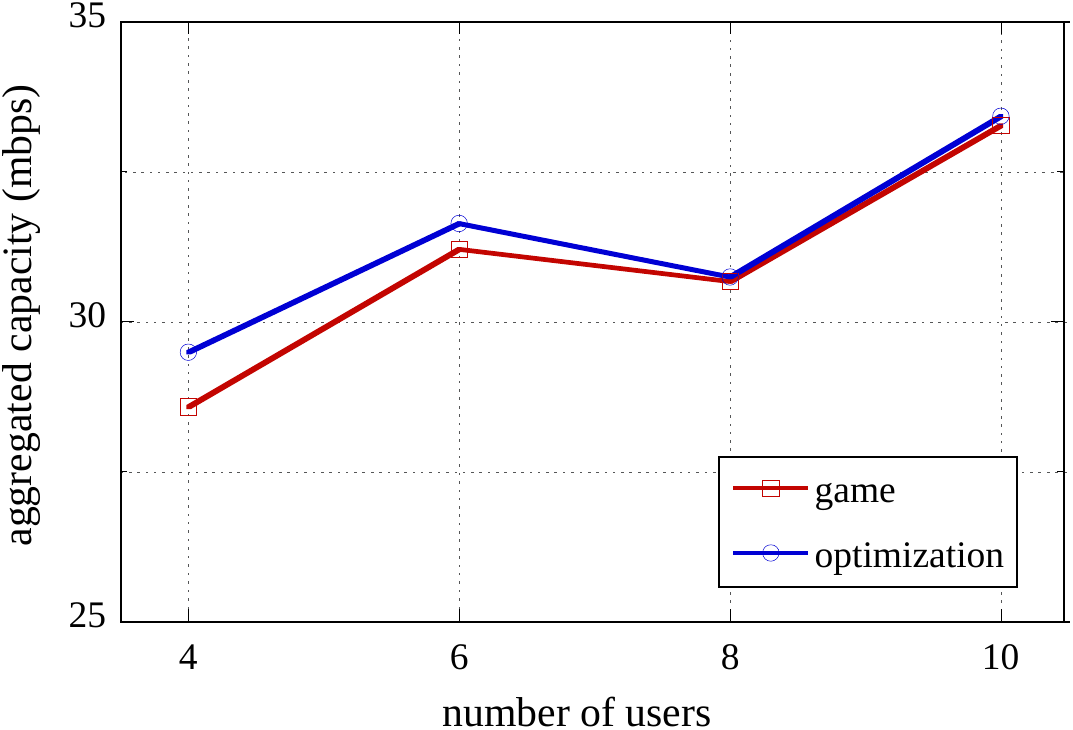}
\caption{Aggregated capacity for the C-Game vs. optimum solution.}
\label{fig:Cap-opt}
\end{figure}

\begin{figure}[h]
\centering
\includegraphics[scale=0.5]{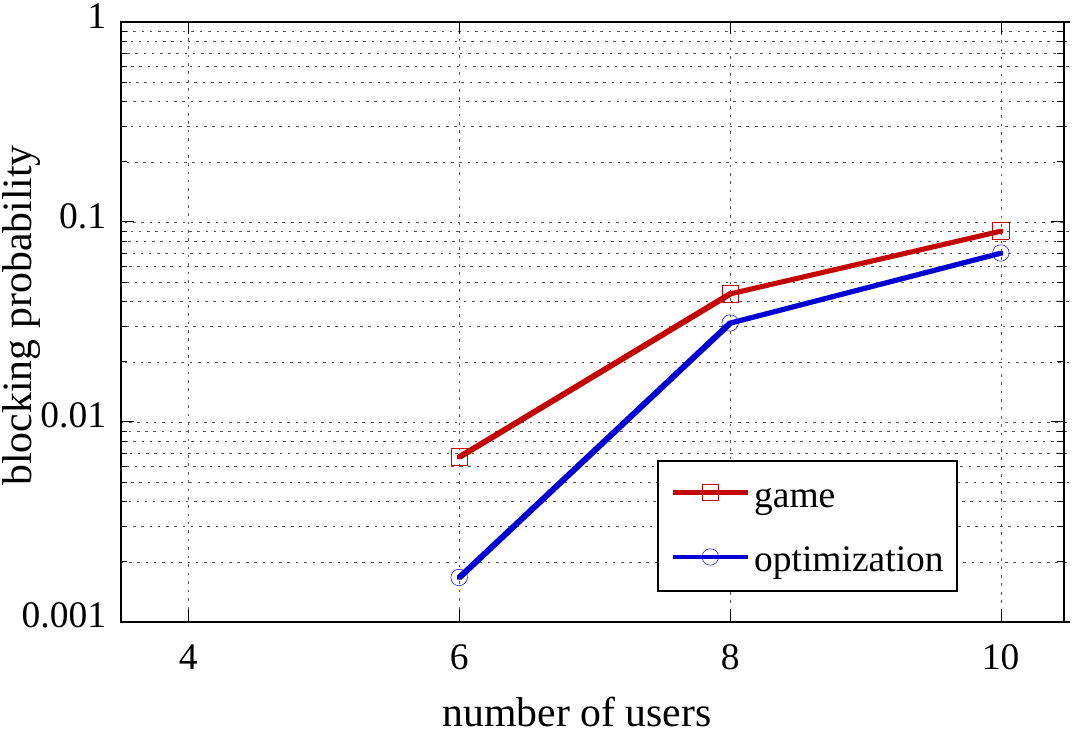}
\caption{Blocking probability for the C-Game vs. optimum solution.}
\label{fig:Block-opt}
\end{figure}

\begin{figure}[h]
\centering
\includegraphics[scale=0.5]{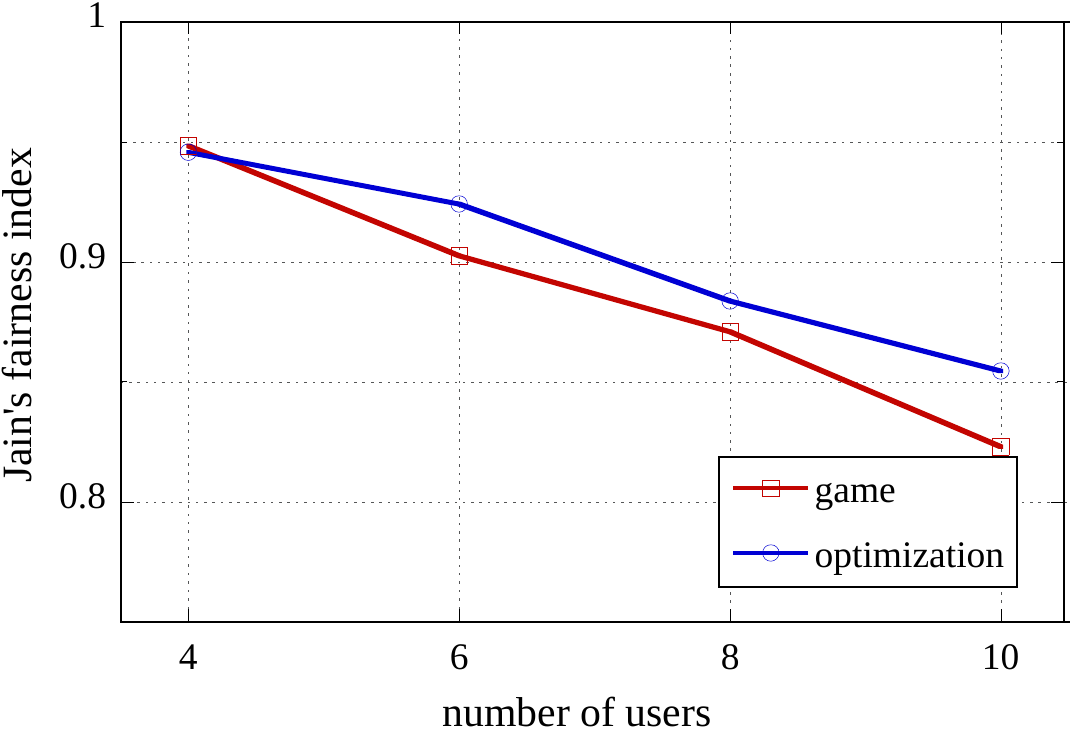}
\caption{Jain's fairness index for the C-Game vs. optimum solution.}
\label{fig:Jain-opt}
\end{figure}

\begin{figure}[h]
\centering
\includegraphics[scale=0.5]{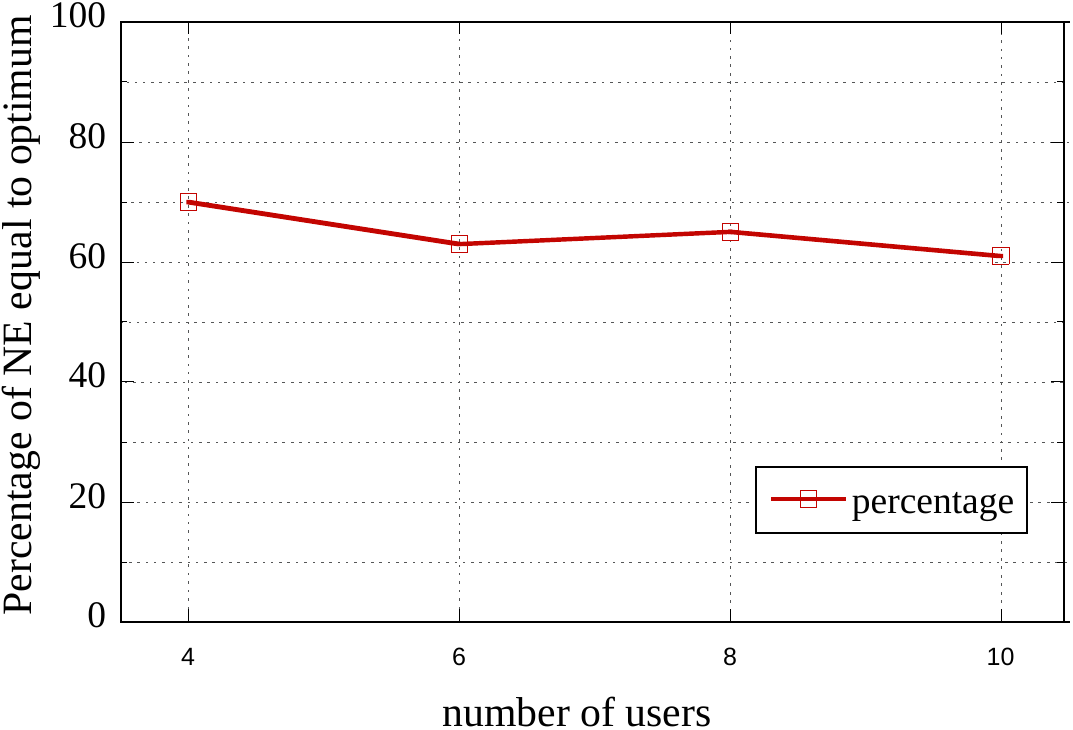}
\caption{Percentage of games (C-Game) reaching a NE equals to the optimum solution.}
\label{fig:Fraction-opt}
\end{figure}

\clearpage
\begin{figure}[h]
\centering
\includegraphics[scale=0.5]{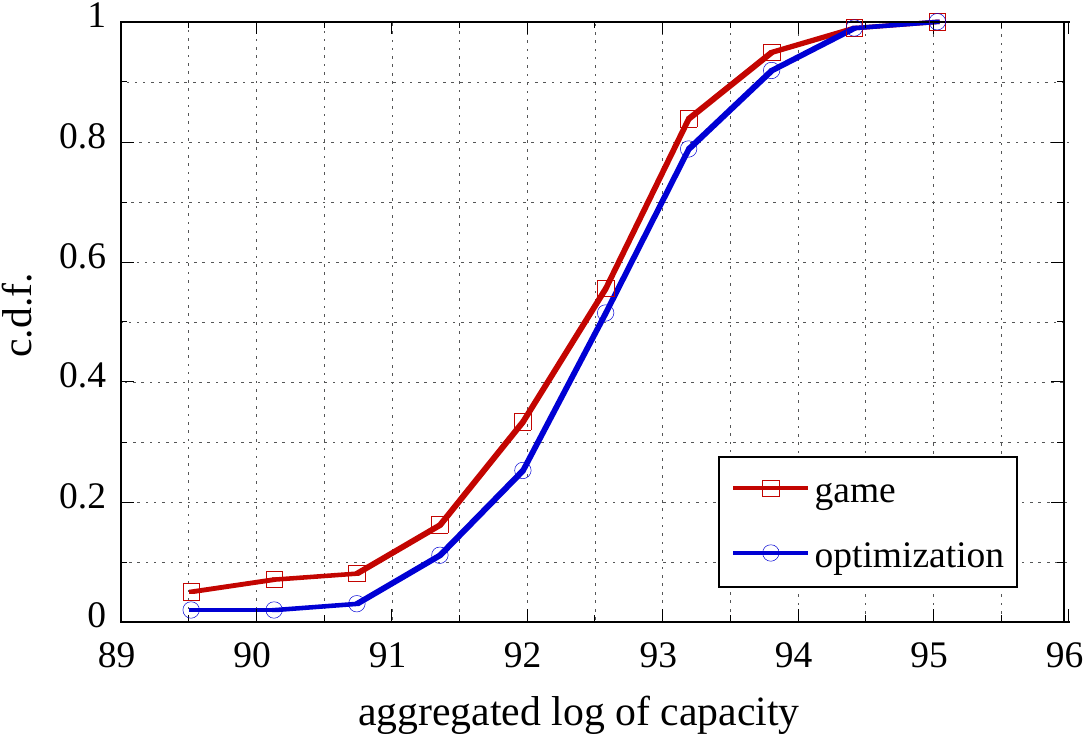}
\caption{Cumulative distribution function of aggregated log-capacity for the C-Game vs. optimum solution.}
\label{fig:Log-opt-cdf}
\end{figure}

\begin{figure}[h]
\centering
\includegraphics[scale=0.5]{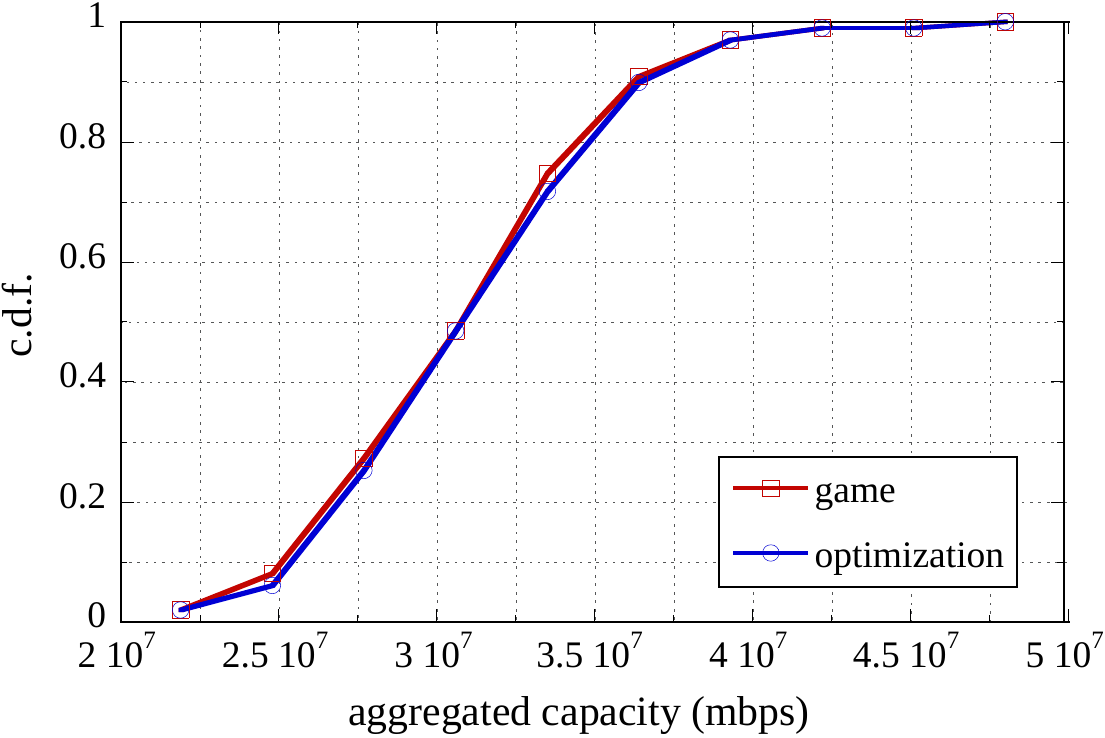}
\caption{Cumulative distribution function of aggregated capacity for the C-Game vs. optimum solution.}
\label{fig:Cap-opt-cdf}
\end{figure}

\begin{figure}[h]
\centering
\includegraphics[scale=0.5]{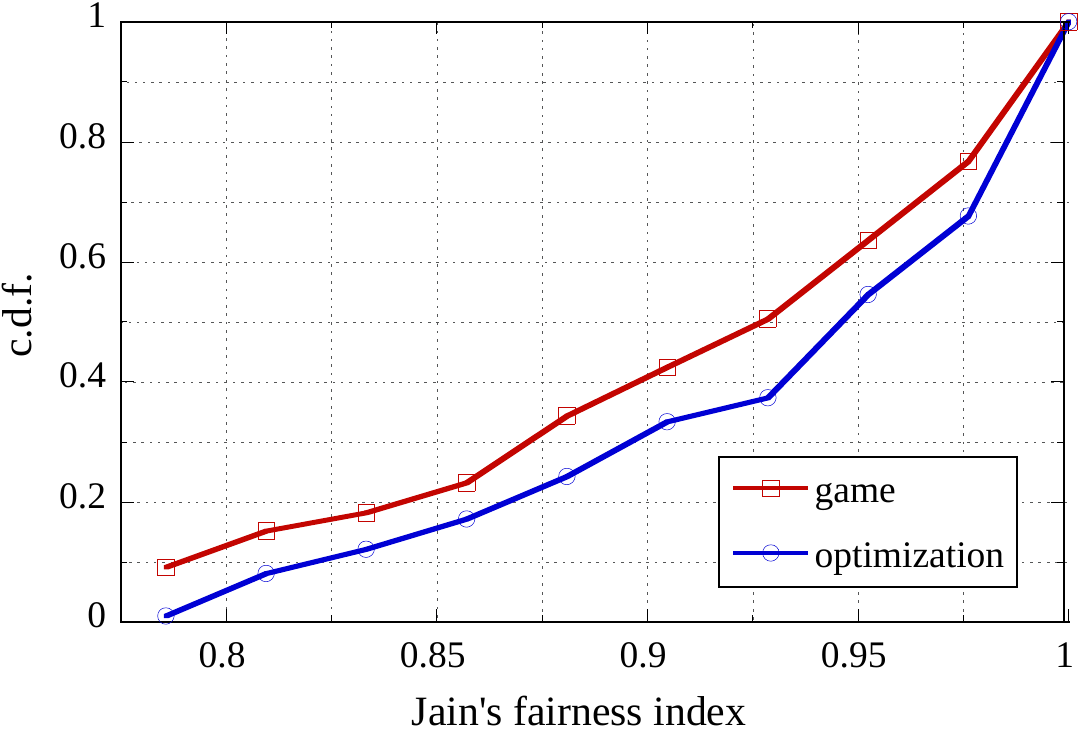}
\caption{Cumulative distribution function of Jain's fairness index for the C-Game vs. optimum solution.}
\label{fig:Jain-opt-cdf}
\end{figure}

Figs.~\ref{fig:Log-opt}, \ref{fig:Cap-opt}, \ref{fig:Block-opt} and \ref{fig:Jain-opt} show the same performance metrics as Figs.~\ref{fig:Log-games} to \ref{fig:Jain-games}. In all the cases, the C-Game achieves an average performance very close to the optimal solution. In addition, Fig~\ref{fig:Fraction-opt} shows the percentage of simulated games that have converged to the optimal solution\footnote{As stated in Section \ref{sec:game}, the optimal solution is always a Nash Equilibrium of a potential game and therefore it can be a possible outcome when the game is played.}. This percentage keeps relatively stable as traffic grows. Finally, Figs.~\ref{fig:Log-opt-cdf}, \ref{fig:Cap-opt-cdf}, and \ref{fig:Jain-opt-cdf} show the cumulative distribution function of the aggregated log-capacity, aggregated capacity and Jain's fairness index in the 100 evaluated random instances for both the game and the optimal solution. It can be seen that the distributions for the game are quite close to the optimum, which shows that performance of the C-Game is always close to the optimum in all the evaluated scenarios. These results confirm the validity of the proposed C-Game to perform distributed optimization with a low computational complexity.

\section{Conclusions}
\label{sec:conclusions}
In this work we have modeled under a game theoretic framework the joint user association and power and channel allocation in a technology-agnostic wireless network with backhaul constraints. Specifically, we have proposed two different potential games denoted as User Game and Channel Game, with different degrees of complexity. Additionally, we have formulated mathematically and solved with the branch-and-bound algorithm the equivalent optimization problem to evaluate the efficiency of these games. Since this problem implies non linear terms and integer variables, its exact resolution is not feasible in real scenarios.

Simulation results have shown that the use of a logarithmic utility function provides a great improvement in the network fairness without excessively reducing the overall network capacity. Additionally, the Channel Game provides a performance almost equal to the much more complex User Game and close to the optimal solution, which suggests its potential application as a distributed resource allocation algorithm to be used in a cloud-based approach.

\section*{Acknowledgments}

This work has been supported by the Spanish Government through the grants TEC2011-23037 and TEC2014-52969-R from the Ministerio de Ciencia e Innovaci\'on (MICINN), Gobierno de Arag\'on (research group T98) and the European Social Fund (ESF), the University of Zaragoza through the grant JIUZ-2013-TEC-07 and the European Project Wi-5 (H2020 Proposal No: 644262).

\section*{References}

\bibliography{mybibfile}

\end{document}